\documentclass[twocolumn]{aa}
\usepackage{graphicx}
\usepackage{natbib}
\bibpunct{(}{)}{;}{a}{}{,}
\usepackage{graphicx}
\usepackage{revsymb}	
\usepackage{amsmath}	
\usepackage[usenames]{color}	
\usepackage{epstopdf}	
\newcommand{\ds}{\displaystyle}
\newcommand{\inv} {\frac {1}}

\newcommand{\derivp} [2] {\frac {\partial #1 } {\partial #2} }
\newcommand{\deriv} [2] {\frac {\textrm{d} #1 } {\textrm{d} #2} }

\newcommand{\eq}[1] {Eq.\,(\ref{#1})}
\newcommand{\eqn} [1] {
\begin{equation} #1
\end{equation}}
\newcommand{\eqna} [1] {
\begin{eqnarray} #1
\end{eqnarray}}

\begin{document}
   \title{Stochastic excitation of non-radial modes}
   \subtitle{I. High-angular-degree $p$~modes}

   \author{K.~Belkacem \inst{1}
           \and
           R.~Samadi \inst{1}
           \and
           M.-J.~Goupil \inst{1}
           \and
           M.-A.~Dupret\inst{1}
           }

 \institute{$^1$Observatoire de Paris, LESIA, CNRS UMR 8109, 92190 Meudon, France}

   \offprints{K. Belkacem}
   \mail{Kevin.Belkacem@obspm.fr}
  \date{}
  \titlerunning{Stochastic excitation of non-radial modes. I.}

  \abstract
    {Turbulent motions in stellar convection zones generate acoustic energy, part of 
    which is then supplied to normal modes of the star.
    Their amplitudes result from a balance between the efficiencies of excitation and damping
     processes in the convection zones. 
  }
    {We develop a formalism that provides the excitation rates 
    of non-radial global modes excited by turbulent convection.
    As a first application, we  estimate the impact of non-radial effects  
     on excitation rates and amplitudes of high-angular-degree modes which are observed on
    the Sun. 
   }
   {A model of stochastic excitation by turbulent convection 
    has been developed to compute the excitation rates, and it has
    been successfully applied to solar radial modes (Samadi \& Goupil 2001,
     Belkacem et al. 2006b). 
     We generalize this approach to the case 
     of non-radial global modes. This enables us to estimate 
    the energy supplied to  high-($\ell$) acoustic modes.
    Qualitative  arguments as well as numerical calculations are 
   used to illustrate the results.
  }
   {We find  that non-radial effects  for $p$~modes are non-negligible: \\
  - for high-$n$ modes (i.e. typically $n > 3$) and for high values of $\ell$; 
    the power supplied to the oscillations depends on the mode inertia.\\
  - for low-$n$~modes, independent of the value of $\ell$, 
  the excitation is dominated by the non-diagonal components of the Reynolds stress term. 
  }
   {We carried out a numerical investigation of high-$\ell$ $p$~modes and we find that 
   the validity of the present formalism is limited to $\ell < 500$ due to the spatial separation of scale 
   assumption. Thus, a model for very high-$\ell$ $p$-mode excitation rates calls for further theoretical 
   developments, however the formalism is valid for solar 
   $g$~modes, which will be investigated in a paper in preparation.

 }   
   \keywords{convection - turbulence - Sun:~oscillations}

   \maketitle
%

\section{Introduction}
\label{intro}
Amplitudes of solar-like oscillations result from a balance between
stochastic excitation and damping in the outermost layers of the
convection zone, which extends to near the surface of the star.
Accurate measurements of the rate at which acoustic energy is supplied
to the solar $p$~modes are available from
ground-based observations (GONG, BiSON) as well as from spacecraft (SOHO/GOLF and
MDI).  From those measurements and a comparison with theoretical
models, it has been possible to demonstrate that excitation is due
to eddy motions in the uppermost part of the convection zone and by
advection of entropy fluctuations.

Stochastic excitation of \emph{radial} modes by turbulent convection has been
investigated by means of  several semi-analytical approaches
\citep{GK77,GK94,B92,Samadi00I}; they differ from each other in the
nature of the assumed  excitation sources, the assumed simplifications
and approximations, and also by the way the turbulent convection is
described \citep[see reviews by][]{Stein04,Houdek06}.  Two major
mechanisms have nevertheless been identified to drive the resonant $p$~modes of the
stellar cavity: the first is related to the Reynolds stress tensor
and as such represents a mechanical source of excitation; the second
is caused by the advection of turbulent fluctuations of entropy by
turbulent motions (the so-called ``entropy source term'') and as it such
represents a thermal source of excitation \citep{GK94,Stein01B}.
Samadi \& Goupil (2001, hereafter Paper~I) proposed a generalized formalism,
taking the Reynolds and entropy fluctuation source terms into
account.  In this model, the source terms are written as functions
of the turbulent kinetic energy spectrum and the temporal-correlation
function. This allows us to investigate several possible models of
turbulence \citep{Samadi02II,Samadi02I}.  The results have been
compared with GOLF data for radial modes, and the
theoretical values are found to be in good agreement with the observations
\citep{Samadi02I}.  Part of the remaining discrepancies has been
recently removed by taking into account the asymmetry introduced by turbulent plumes
\citep{Belkacem06a,Belkacem06b}.

In this paper we take an additional step in extending the
\cite{Samadi00I} formalism to the case of non-radial global modes.
This will enable us to estimate the excitation rates for a wide variety of
$p$ and $g$ modes excited in different types of stars.  The present
model provides the energy supplied to the modes by turbulence in inner,
as well as outer, stellar convective regions, provided the turbulent
model appropriate for the relevant region is used.  Studies of
stochastic excitation of solar radial modes
\citep{Samadi02II,Samadi02I} have given us access mainly to the radial
properties of turbulence. The present generalized formalism 
enables us to take into account the horizontal properties of turbulence (through the 
non-radial components of the Reynolds stress) in the
outermost part of the convective zone.

In the Sun, high-angular-degree $p$~modes (as high as one thousand) have
been detected \citep[e.g.,][]{Rabello04}. 
From an observational point of view,
\cite{Woodard01} found that the energy supplied
to the mode increases with $\ell$, but that above some high-$\ell$
value, which depend on the radial order $n$
\citep[see][Fig.~2]{Woodard01}, the energy decreases with increasing
$\ell$. They mentioned the possibility of an unmodelled mechanism of
damping. 
Hence one of the motivations of this work is to investigate such an issue. 
As a first step,  we develop here a theoretical model of the stochastic excitation taking into 
account the $\ell$-dependence of the source terms 
to seek a physical meaning for such a behaviour of the amplitudes. 

Modelling of the mechanisms responsible for the excitation of
non-radial modes is not only useful for high-$\ell$ acoustic modes but
also for gravity modes, which are intrinsically non-radial. As for $p$ modes, $g$ modes are stochastically
excited by turbulent convection; the main difference is that the dominant
restoring force for $g$ modes is buoyancy. We however stress that convective 
penetration is another possible excitation mechanism for $g$~modes \citep[e.g.][]{Dintrans05}.
Such modes are
trapped in the radiative interior of the Sun, so their detection promises
to give a much better knowledge of the deep solar interior.  However, such
modes are evanescent in the convection zone; their amplitudes at the
surface are very small and their detection remains controversial.  A
theoretical prediction of their amplitudes is thus an important issue.
It requires an estimation of the excitation rates but also  of the
damping rates. Unlike $p$~modes, the damping rates cannot be inferred
from observations and this introduces considerable uncertainties; e.g.
theoretical estimates of the $g$-mode amplitudes
\citep{Gough85,Kumar96} differ from each other by orders of magnitudes,
as pointed out by \cite{CD2002}.  We thus stress that the present  work
focuses on the excitation rates -- damping rates are not investigated.  A
specific study of gravity modes will be considered in a forthcoming
paper.

The paper is organized as follows: Sect.~2 introduces the general
formalism, and a detailed derivation of the Reynolds and entropy source
terms is provided.  In Sect.~3, we demonstrate that the formalism of
\cite{Samadi00I} is a special case and an asymptotic limit  of the
present model.  In Sect.~4, we determine, using qualitative arguments,
the different contributions to the excitation rates and identify the
dominant terms involving  the angular degree (\,$\ell$\,).  Sect.~5 presents the
numerical results: excitation rates are presented and discussed.
Sect.~6 discusses the limitations of the model and some conclusions
are formulated in Sect.~7.


\section{General formulation}
\label{amplitude}

Following Paper~I, we start from the perturbed
momentum and continuity equation
\begin{eqnarray}
\label{momentum}
& &\derivp{(\rho_0 + \rho_1) \vec v}{t} + \nabla : \left(\rho_0 \vec v \vec v \right) = \nonumber \\ 
& &\hspace{3cm} -\nabla p_1
+ \rho_1 \vec g_0 + \rho_0 \vec g_1 + \rho_1 \vec g_1 \; ,
\end{eqnarray}
\begin{eqnarray}
\label{continuity}
\derivp{\rho_1}{t} + \nabla \cdot \left( (\rho_0+\rho_1) \vec v \right) = 0
\end{eqnarray}
where $\rho$ is the density, $p$ is the
pressure,  and $\vec g$ is the gravity. 
The subscripts 1 and 0 denotes Eulerian perturbations and
equilibrium quantities respectively,  except for velocity where the subscript
1 has been dropped for ease of notation. In the following, the
velocity field is split into two contributions, namely the
oscillation velocity ($\vec v_{osc}$) and the turbulent velocity field
($\vec u$), such
that $\vec v= \vec v_{osc} + \vec u$. For a given mode, the
 fluid displacement can be written as:
\begin{equation}
\label{displacement_trapping}
\vec {\delta r_{osc}} = \frac{1}{2} \left( A(t) \, \vec \xi(\vec r) e^{-i\omega_0 t} + c.c. \right) \; ,
\end{equation}
where $\omega_0$ is the eigenfrequency, $\vec \xi (\vec r)$ is the displacement  eigenfunction
 in the absence of turbulence, $A(t)$
is the amplitude due to the turbulent forcing, and $c.c$ denotes the complex conjugate.\\
The power (\,$P$\,) injected into the modes is  related to  the mean-squared
 amplitude (\,$<|A|^2>$\,)  by (see Paper~I)
\begin{equation}
\label{puissance_trapping}
P = \eta <|A|^2> I ~\omega_0^2 \; ,
\end{equation}
where the operator $<>$ denotes a statistical average performed
on an infinite number of
independent realizations, $\eta$ is the damping rate, and $I$ is the mode inertia.

We use the temporal WKB assumption, i.e. that $A(t)$ is slowly varying with respect to the oscillation period, 
$\eta \approx \textrm{d ln} A(t) / \textrm{d}t \ll \omega_0$ (see Paper~I for details). \\
Under this assumption, using \eq{displacement_trapping} with \eq{momentum} and \eq{continuity} 
(see Paper~I) yields:
\begin{equation}
\label{equation_amplitude_trapping}
\frac{\textrm{d} A(t)}{\textrm{d}t} + \eta A(t) = \frac{1}{2\omega^2_0 I} \int \textrm{d}^3x
\  {\vec \xi^\ast} \cdot \derivp{{\vec S}}{t} e^{i\omega_0t}\; ,
\end{equation}
where $\textrm{d}^3 x$ is the volume element 
and $\vec S=-(\vec f_t + \vec \nabla h_t + \vec g_t)$ the excitation source
terms. Temporal derivatives appearing in \eq{equation_amplitude_trapping} are
\begin{itemize}
\item the Reynolds stress contribution:
\begin{equation}
\label{Rey}
\derivp{f_t}{t}= - \derivp{}{t} \left(\nabla : (\rho_0 \vec u \vec u)\right) \, ,
\end{equation}
where $\vec u$ is the turbulent component of the velocity field.
\item The entropy term:
\begin{equation}
\label{entrop}
\derivp{}{t} \nabla h_t= -\nabla \left( \alpha_s \, \frac{ \textrm{d} {\delta s_t}}{\textrm{d}t}  - \alpha_s  \vec u \cdot
\nabla \vec s_t \right) \, ,
\end{equation}
where $\delta s_t$ is the turbulent Lagrangian fluctuation of the entropy (\,$\alpha_s=
\textrm{d} p_1 / \textrm{d} s_t$\,) and $p_1$ denotes the Eulerian pressure fluctuations.\\
This contribution represents the advection of entropy fluctuations by turbulent motion and as such 
is a thermal driving. Note that it has been shown by \cite{Belkacem06b} that this term is necessary 
to reproduce the maximum in the amplitude as a function of frequency in the case of solar radial $p$~modes. 
\item the fluctuating gravity term
\begin{equation}
\label{buo}
\derivp{g_t}{t}=\derivp{\rho_1 g_1}{t} \, .
\end{equation}
where $g_1$ is the fluctuation of gravity due to the turbulent field. 
This contribution  can be shown to be negligible and will not be considered in detail here for $p$ modes. 
\end{itemize}
Several other excitation source terms appear in the right-hand side of \eq{momentum}.  
However, as shown in Paper~I, their 
contributions are  negligible since 
they are linear in terms of turbulent
fluctuations.\footnote{Linear terms are defined as the product of an equilibrium quantity and a fluctuating one.}

From \eq{equation_amplitude_trapping}, one obtains the mean-squared amplitude
\begin{eqnarray}
<\left| A \right|^2 (t)> &=& \frac{e^{-2\eta t}}{4(\omega_0I)^2} \int_{-\infty}^{t} \textrm{d}t_1 \textrm{d}t_2 \nonumber \\
 & &\hspace{-0.7cm}  \times  \int \textrm{d}^3r_1 \textrm{d}^3r_2  e^{\eta(t_1+t_2)+i\omega_0(t_1-t_2)} \nonumber \\
& & \hspace{-0.7cm} \times \left< \left(\vec \xi^\ast(\vec r_1) \cdot \vec S(\vec r_1, t_1) \right)\left(\vec
\xi(\vec r_2) \cdot \vec S^\ast(\vec r_2, t_2) \right)\right>  \, ,
\end{eqnarray}
where the subscripts $1$ and $2$ denote two spatial and temporal locations. \\
To proceed further, it is convenient to define the following coordinates:
\eqn { \begin{array} {lcr}
\vspace{0.2cm} \vec x_0 =\displaystyle {  \frac {\vec r_2 + \vec r_1} {2}}  & &
t_0
= \displaystyle { \frac {t_1 +t_2} {2} } \nonumber \\
\vec r = \vec r_2 - \vec r_1 & & \tau = t_2 - t_1 \nonumber
\end{array}
}
where $\vec x_0$  and $t_0$ are the average space-time position  and $\vec r $ and $\tau $ are
 related to the local turbulence.

In the following, $\vec{\nabla}_0 $ is the large-scale derivative associated
with  $\vec{ x}_0 $,  $\vec{\nabla}_{\vec r} $ is the small-scale one associated with
$\vec{r}$ and the derivative
operators $\vec{\nabla}_1$ and $\vec{\nabla}_2 $ are associated
with  $\vec r_1$ and $\vec r_2$ respectively.
The mean-squared amplitude  can be rewritten in terms
of the new coordinates as
\eqna{
\left <  \mid A \mid ^2 (t) \right > &  = & \inv{4 (\omega_0 I)^2} \nonumber \\
& & \hspace{-2.2cm} \times \int_{-\infty}^{t} \textrm{d}t_0 \, e^{2 \eta (t_0-t)}
\int_{2 (t_0-t)}^{2(t-t_0)} \textrm{d}\tau    
 \int  \textrm{d}^3x_0 \, \textrm{d}^3r \, \, e^{-i\omega_0 \tau}  \nonumber \\
& & \hspace{-2.2cm} \times  \left <  \left (   \vec \xi^*    \cdot    \vec {\cal S}  [ \vec x_0-
\frac{\vec r}{2}, t_0 - \frac{\tau} {2}] \right ) \left (
      \vec \xi   \cdot    \vec {\cal S}^* [\vec x_0+\frac{\vec r}{2}, t_0
+ \frac{\tau} {2}]  \right )    \right > \; .
\label{eqn:A2t_2}
}
Subscripts   1 and 2 are the  values taken at the spatial and  temporal
positions $[ \vec x_0-\frac{\vec r}{2}, - \frac{\tau} {2}]$ and $ [\vec
x_0+\frac{\vec r}{2}, \frac{\tau} {2}]$ respectively.  In the
excitation region, the eddy lifetime  is much smaller than the
oscillation lifetime ($\sim 1/\eta$) of $p$ modes such that the
integration over $\tau$ can be extended   to infinity. Hence all time
integrations over $\tau$ are understood
to be performed  over the range  $]-\infty , +\infty [$.

We assume  a stationary  turbulence, therefore the source
term (\,$S$\,) in Eq.\,(\ref{eqn:A2t_2}) is invariant to translation in $t_0$.
Integration over $t_0$ in Eq.\,(\ref{eqn:A2t_2}) and using the definition of $\vec S$ 
(\eq{Rey}, \eq{entrop} and \eq{buo}) yields:
\begin{equation}
\label{mean_square_amplitude_trapping}
<|A|^2> = \frac{1}{8 \eta (\omega_0 I)^2} \left( C_R^2 + C_S^2 + C_{RS} \right) \;,
\end{equation}
where $C_R^2$ and $C_S^2$ are the turbulent Reynolds
 stress and entropy fluctuation contributions whose expressions are respectively:
\begin{itemize}
\item the Reynolds source term:
\eqna{
C_R^2 & =& \int  \textrm{d}^3x_0   \,  \int_{-\infty}^{+\infty} \textrm{d}\tau \, e^{-i\omega_0 \tau}
\int \textrm{d}^3r  \nonumber \\  & & \times \left < \,
\left (\rho_0 ~u_j u_i ~ \nabla_0^j  \xi^i \right )'
\left (\rho_0 ~u_l u_m ~\nabla_0^l  \xi^{*m} \right )''
\right >
\label{eqn:C2R}
}
where a {\it separation of scales} is assumed, i.e. that 
the spatial variation of the eigenfunctions is large compared to the typical length scale of turbulence 
(see Sect.~\ref{discussion} for a detailed discussion).
\item the entropy contribution
\eqna{
C_S^2 & =  &   \int  \textrm{d}^3x_0   \,  \int_{-\infty}^{+\infty} \textrm{d} \tau \, e^{-i\omega_0 \tau}
\int \textrm{d}^3r   \nonumber \\
& &\times  \left < \,
 \left ( h_{t} \, \nabla_{0j} \xi^j      \right)_1
\left (  h_{t} \,   \nabla_{0l}  \xi^{*l}    \right )_2  \,
 \right >
\label{eqn:C2S}
}
\end{itemize}
$C_{RS}$ is the cross-source term,  
it represents interference between source terms. 
For $p$~modes, $C_{RS}$ turn out to be negligible 
because it involves third-order correlation products which are small 
 and strictly vanish under the QNA assumption \citep{Belkacem06b}.  

\subsection{Turbulent Reynolds stress contribution}
\label{reynolds}

\eq{eqn:C2R} is first rewritten as: 
\eqna{
C_R^2 & =& \int  \textrm{d}^3x_0   \,  \int_{-\infty}^{+\infty} \textrm{d}\tau \, e^{-i\omega_0 \tau}
\int \textrm{d}^3r \, \rho_0^2  \times \nonumber \\  & &  \nabla_0^j  \xi^i \left < \,
  \left (u_j u_i  \right )_1
\left (u_l u_m\right )_2   
\right > \nabla_0^l  \xi^{*m}
\label{eqn:C2Rb}
}
The fourth-order moment is then  approximated assuming the  quasi-normal
approximation (QNA, \citep[][Chap. VII-2]{Lesieur97}) as in Paper~I. The QNA is 
  a convenient means 
to decompose the fourth-order velocity correlations in terms of a product 
of second-order velocity correlations; that is, one uses
\begin{eqnarray}
\label{fourth-order}
 \langle (u_i u_j)_{(1)} ({u_l u_m})_{(2)} \rangle  &=&  
  \langle (u_i u_j)_{(1)}   \rangle \, \langle (u_l u_m)_{(2)}  \rangle \nonumber \\
& & \hspace{-0.5cm} +  \langle (u_i)_{(1)}  (u_l)_{(2)}  \rangle \, \langle (u_j)_{(1)}  (u_m)_{(2)} 
\rangle  \nonumber \\
 & &\hspace{-0.5cm} + \langle (u_i)_{(1)}  (u_m)_{(2)}  \rangle \, \langle (u_j)_{(1)}  (u_l)_{(2)}  \rangle  
\end{eqnarray}
A better approximation is the closure model with plumes
\citep{Belkacem06a,Belkacem06b} which can be adapted to the present
formalism in order to take into account the presence of up and
downdrafts in the solar convection zone. 

It then is possible to
express the Fourier transform of the resulting second-order moments
in term of the turbulent kinetic and entropy energy spectrum
(see Paper~I for details)
\begin{equation}
\phi_{ij} = TF(\langle u_i u_j  \rangle) = \frac{E(k,\omega)}{4\pi k^2} \left ( \delta_{ij} - \frac{k_i k_j}{k^2} \right ) \, ,
\end{equation}
where $E(k,\omega)$ is the turbulent kinetic energy spectrum.

 The turbulent
Reynolds term  \eq{eqn:C2R} takes the following general expression under the
assumption of isotropic turbulence:
\begin{eqnarray}
\label{C2R_2}
C_R^2 & =  &\pi^{2} \int  \textrm{d}^3 x_0  \,
\left (\rho_0^2 \,  b^*_{ij} b_{lm} \right )
{S}_{(R)}^{ijlm}(\omega_0)
\end{eqnarray}
where
\begin{eqnarray}
\label{SR_trapping}
{S}_{(R)}^{ijlm} &=&  \int_{-\infty}^{+\infty} \textrm{d}\omega \int \textrm{d}^3k \,
\left (  T^{ijlm} + T^{ijml} \right ) \nonumber \\
&\times& \frac {E^2(k)} {k^4 }  \chi_k( \omega_0 + \omega) \, \chi_k(\omega) \\
T^{ijlm} &=& \left( \delta^{il}- \frac {k^i k^l} {k^2}  \right)
\left( \delta^{jm}- \frac {k^j k^m} {k^2}  \right) \\
\label{b_ij}
b_{ij} &\equiv& \vec {e}_i \, \cdot \, \left ( \vec \nabla_0 \, : \, \vec \xi \right )
\, \cdot \,  \vec {e}_j
\end{eqnarray}
where $\{\vec e_i\}$ are the spherical coordinate unit vectors, $(\vec k,\omega)$ 
are the wavenumber and frequency of the
turbulent eddies and turbulent kinetic energy spectrum $E(\vec k,
\omega)$, which is expressed as the product $E(\vec k) \,
\chi_k(\omega)$  for isotropic turbulence \citep{Lesieur97}. 
 The kinetic energy spectrum $E(k)$ is   normalized as
\eqn{  
\int_0^{\infty}  dk \, E(k)  = \frac{1}{2} \,  \Phi w^2 
\label{eqn:E:normalisation}
}
where $w$ is an estimate for the vertical convective velocity  
and  $\Phi$ is a factor  introduced by  \citet{Gough77} to take into account
anisotropy effects. 
A detailed
discussion of the temporal correlation function ($\chi_k$) is addressed in \citet{Samadi02I}. \\

The contribution of the Reynolds stress can thus be written as
 (see Appendix~\ref{C_R_detail}):
\begin{eqnarray}
\label{C2R_ref}
C_R^2 & =  &  4\pi^{3}   \int  \textrm{d}m   \,  \; R(r)~ S_R(\omega_0)\; ,
 \end{eqnarray}
with
\begin{eqnarray}
\label{gammabeta}
R (r) &=&   {16\over 15} ~  \left| \deriv{\xi_r}{r}  \right|^2  +  {44\over 15} ~    \left| \frac{\xi_r}{r}  \right|^2
+   \frac{4}{5} \left( \frac{\xi^*_r}{r} \deriv{\xi_r}{r} + c.c \right) \nonumber \\ 
&+&  ~ L^2 \left( {11\over 15} ~  \left| \zeta_r  \right|^2 - {22 \over 15} (\frac{\xi_r^* \xi_h}{r^2} +c.c) \right) \nonumber \\
 &-&{2\over 5} L^2 \left(\deriv{\xi^*_r}{r} {\xi_h \over r} 
+ c.c \right)\nonumber \\
&+&  \left| \frac{\xi_h}{r}  \right|^2 
\left( \frac{16}{15} L^4+ \frac{8}{5} {\cal F}_{\ell,\vert m \vert} - \frac{2}{3} L^2 \right) 
\end{eqnarray}
where we have defined
\begin{eqnarray}
L^2 &=& \ell (\ell + 1) \\
\zeta_r &\equiv&  \deriv{\xi_h}{r} +\frac{1}{r}(\xi_r-\xi_h) \\
{\cal F}_{\ell,|m|} &=&  \frac{|m| (2 \ell + 1)}{2} \left( \ell(\ell+1) - (m^2 + 1 )\right) \\
S_R(\omega_0) &=& \,\int  \frac {\textrm{d}k} {k^2 }~E^2(k) ~\int \textrm{d}\omega 
~\chi_k( \omega + \omega_0) ~\chi_k( \omega )
 \end{eqnarray}
 
 Note that in the present work, nonradial effects are taken into account only  through \eq{gammabeta}. 
A more complete description would require including anisotropic turbulence effects in \eq{SR_trapping}, 
this is however beyond the scope of the present paper. 
 
\subsection{Entropy fluctuations contribution}
\label{C_S^2}
The entropy source term is computed as for the Reynolds contribution 
 in Sect.~\ref{reynolds}. 
Then \eq{eqn:C2S} becomes
\begin{equation}
\label{C2S_62}
C_S^2  = \frac{2 \pi ^2}{\omega_0^2}  \, \int \textrm{d}^3 x_0 \,   \alpha_s^2
 \, h^{ij} \, {S}^{(S)}_{ij} (\omega_0) \; ,
\end{equation}
where
\begin{eqnarray}
\label{SS_trapping}
& & {S}^{S}_{ij} (\omega_0)=
\int \textrm{d}^3k \,  T_{ij} \,  \frac { E( k) } {  k^2} \, \frac {E_s (k) } { k ^ 2} \,
\int \textrm{d}\omega \,\chi_k ( \omega_0+ \omega ) \chi_k ( \omega ) \nonumber \\
& & {\rm with} \nonumber \\
& & T_{ij} = \left( \delta_{ij}- \frac {k_i k_j} {k^2}  \right)
\end{eqnarray}
where $ E_s(k)$ is the entropy spectrum (see Paper~I),
and
\begin{eqnarray}
\label{hij}
h^{ij} & =&  \left | \mathcal{C} \right |^2 \, \nabla_1^i( \ln \mid \alpha_s \mid) \,
\nabla_2^j( \ln \mid \alpha_s
\mid ) \nonumber \\
& & -  \mathcal{C}^\ast \,  \nabla_1^i( \ln \mid \alpha_s \mid) \, \nabla_2^j  ( \mathcal{C}  ) \nonumber  \\
& & - \mathcal{C}  \,  \nabla_1^i( \ln \mid \alpha_s \mid) \, \nabla_2^j(  \mathcal{C}^\ast )
 + \nabla_1^i( \mathcal{C}^\ast  ) \, \nabla_2^j(  \mathcal{C}  )  \; ,
\end{eqnarray}
where $\mathcal{C} \equiv  \nabla . \, \vec \xi $ is the mode compressibility.

The final expression for the contribution of entropy fluctuations 
reduces to  (see Appendix~\ref{C_S_detail} ):
\eqn{
\label{C_S_ref}
C_S^2  = \frac{4 \pi^3 \, \mathcal{H}}{\omega_0^2}  \, 
\int \textrm{d}^3 x_0 \, \alpha_s^2 \, \left ( A_\ell + B_\ell \right ) \,
\mathcal{S}_S(\omega_0)
}
where $\mathcal{H}$ is the anisotropy factor introduced in Paper~I which,
 in the current
assumption (isotropic turbulence),  is equal to  $ 4/3 $.  In addition,
\eqna{
\label{C_S_ref2}
A_\ell & \equiv & \frac{1 }{r^2}  \,  \left|  D_\ell \, 
\deriv{\left( \ln \mid \alpha_s \mid \right)}{\ln r}
 - \deriv{D_\ell }{\ln r}  \right|  ^2    \label{eqn:Al}
\\
\label{Bell}
B_\ell & \equiv &  \frac{1 }{r^2} \, L^2 \, \left| D_\ell \right| ^2
\label{eqn:Bl}
\\
\mathcal{S}_S(\omega_0) & \equiv & \int \frac{\textrm{d}k}{k^4}\,E(k)
\, E_s(k) \, \int \textrm{d}\omega \,
\chi_k(\omega_0+\omega)\,  \chi_k(\omega)\label{eqn:FS} } where
\eqna{ \label{C_S_ref3}
 D_\ell(r,\ell) \equiv  D_r - \frac{L^2 } {r} \, \xi_h \; , \;
D_r \equiv \frac{1}{r^2} \, \deriv{}{r} \left (  r^2  \xi_r \right ) \; .
}
\section{The radial case}
\label{radial}
We show in this section that we recover the results of Paper~I 
providing that:
\begin{itemize}
\item we restrict ourselves to the radial case by setting $\ell = 0$  ($\xi_h = 0$).
\item we assume a plane-parallel atmosphere.
\end{itemize}
 In  the entropy source term (\,$C_S^2$\,) the mode compressibility  for a radial mode
becomes:
\begin{equation}
\mathcal{C} = -\frac{\delta \rho}{\rho} = \frac{1}{r^2} \deriv{ \left(r^2 \, \xi_r \right)}{r} ~Y_{\ell,m}
\end{equation}
and from \eq{C_S_ref2} and \eq{Bell}, one then has
\begin{eqnarray}
& & A_{\ell = 0}  =  \frac{1} {r^2} \, \left| D_r \deriv{} {\ln r}   \ln \mid
\alpha_s \mid  - \deriv{D_r} {\ln r}   \right| ^2 \\
& & B_{\ell=0} = 0
\end{eqnarray}
We thus obtain (\eq{C_S_ref})
\begin{eqnarray}
\label{CSrad}
C_S^2  &=& \frac{4 \pi ^3\, \mathcal{H}}{\omega_0^2}  \, \int \textrm{d}^3 x_0 \, \alpha_s^2 \nonumber \\
 &\times&\frac{1} {r^2} \, \left| D_r
\deriv{\ln \mid \alpha_s \mid} {\ln r}     - \deriv{ D_r } {\ln r}\right| ^2
\, \mathcal{S}_S(\omega_0)
\end{eqnarray}

For the Reynolds stress contribution, \eq{C2R_ref} 
reduces to
\begin{eqnarray}
\label{CRrad}
C_R^2 &=&  4 \pi^{3}   \int  d^3x_0   \,  \rho_0^2 \, \nonumber \\
& & \hspace{-1.0cm} \times 
\left( 
\frac{16}{15} \left| \deriv{\xi_r}{r} \right|^2 + \frac{44}{15} \left| \frac{ \xi_r}{r} \right|^2  
      + \frac{4}{5} \frac{ \xi_r^*}{r} \, \deriv{\xi_r}{r} + c.c. 
\right)  
\, \mathcal{S}_R(\omega_0)
\end{eqnarray}
To proceed further, we use the plane-parallel approximation. 
It is justified (for $p$~modes) by the fact that excitation takes place 
in the uppermost part of the convection zone (\,$r / R \approx 1$\,). 
It is  valid when the condition $ r \,  k_{osc} \gg 1$  is  fulfilled
in the excitation region ($k_{osc}$ being the local wavenumber), i.e. where 
excitation is dominant.  
Consequently:
\begin{equation}
\label{pp}
  \ds { \left|  \deriv{ \xi_r}{r} \right| \gg \left| \frac{ \xi_r}{r} \right| } \cdot
\end{equation}
Validity of this inequality has been numerically verified and is discussed in Sect.~\ref{pmodes} (\eq{rapp_dxir})

With \eq{pp}, \eq{CSrad}, and \eq{CRrad} simplify   as:
\begin{eqnarray}
\label{rad}
C_S^2  &=&  \frac{4 \pi ^3\, \mathcal{H}}{\omega_0^2}  \, \int \textrm{d}^3x_0 \, \alpha_s^2  \nonumber \\
 &\times&\frac{1} {r^2} \, \left| \deriv{ \xi_r}{r}
\deriv{\ln \mid \alpha_s \mid} {\ln r}     - \deriv{} {\ln r}
\left(\deriv{\xi_r}{r}\right)  \right| ^2
\, \mathcal{S}_S(\omega_0) \\
C_R^2  &=&  \frac{64}{15} \pi^{3}   \int  \textrm{d}^3 x_0   \,  \rho_0^2 \,   
\left| \deriv{ \xi_r}{r} \right| ^2 \, \mathcal{S}_R(\omega_0)
\end{eqnarray}
These are the expressions obtained by Paper~I and \citet{Samadi05c} for
the radial modes in a plane-parallel geometry.\\

\section{Horizontal effects on the Reynolds and entropy source terms}
\label{pmodes}

We derive asymptotic expressions for the excitation source
terms (\eq{C2R_ref} and \eq{C_S_ref}) in order to identify
the major nonradial contributors to the excitation rates
in the solar case.

\subsection{$\ell$ dependence of the eigenfunctions }
\label{discussion_fct}

$P$ modes are mainly excited in a thin surface layer which is the
super-adiabatic zone at the top of the convection zone.

Let us consider  the equation of continuity and the transverse  
component of the
equation of motion for the oscillations. Let us neglect the Lagrangian  
pressure
variation and Eulerian gravitational potential variation at
$r=R$ (the surface). The ratio of the horizontal to the vertical
displacement at the surface boundary is then  approximately given by \citep[][p. 105]{Unno89}:
\eqna{
\label{rapp_fct}
\frac{ \xi_h}{ \xi_r} \simeq \sigma^{-2} \; ,
}
where $\sigma$ is the dimensionless frequency defined by:
\eqna{
\label{sigma}
\sigma^2 = \frac{R^3}{GM} \omega^2 \; ,
}
where $\omega$ is the angular frequency of the mode, $R$ the star  
radius and $M$ its mass.\\
Frequencies of solar $p$~modes then range between $\sigma \approx 10$ and $\sigma
\approx 50$ ($\nu \in [1 , 5]$ mHz).
Hence, for the solar oscillations,  one always has:
\begin{equation}
\left| \xi_r \right| \gg \left| \xi_h \right|
\end{equation}
However  \eq{gammabeta} and \eq{C_S_ref3} involve coefficients  
depending on the angular degree ($\ell$). We then also consider the ratio
\begin{equation}
\label{rapp_bis}
L^2 \frac{ \xi_h}{\xi_r} \approx L^2~ \sigma^{-2}
\end{equation}
Equation.~(\ref{rapp_bis}) is of order of unity for $\ell \sim \sigma$.
For example, for a typical frequency of $3$~mHz, one can not neglect
the horizontal effect $L^2 \left|\frac{\xi_h}{r}\right|$ in
front of $\left|\frac{\xi_r}{r}\right|$ for values of $\ell$ equal or
larger than 30.

In what follows, we introduce the complex number $f$, 
which is the degree of non-adiabaticity, defined by the relation:
\begin{equation}
f = \frac{\delta p/p}{\Gamma_1 \delta\rho /\rho}.
\end{equation}
Note that $f=1$ for adiabatic oscillations.

Let now  compare the derivatives. Under the same assumptions as above,
neglecting the
term in $(p/\rho) {\rm d}(\delta p/p)/{\rm d}r$ in the radial  
component of the
equation of motion (standard mechanical boundary condition),
one gets near the surface:
\begin{eqnarray}
\label{rapp_dxir}
\frac{{\rm d}\xi_r}{{\rm d}r} / \left(\frac{\xi_r}{r}\right) &\simeq&
(f\Gamma_1)^{-1}
[\sigma^2+2+(L^2/\sigma^2-2)(f\Gamma_1-1)] \nonumber \\
&\simeq& \sigma^2/(f\Gamma_1).
\end{eqnarray}
Hence, we always have 
  $\left| \partial \xi_r / \partial r \right| \gg \left| \xi_r / r \right|$
in the excitation region (except near a node of $\partial \xi_r / \partial r$).
Similarly to \eq{rapp_fct}, one can assume:
\eqna{
\label{rapp_dxihdxir}
\deriv{\xi_h}{r} / \deriv{\xi_r}{r} \simeq \sigma^{-2} \;.
}
In fact, comparing \eq{rapp_dxihdxir} with the numerically-computed
eigenfunctions shows that it holds even better than \eq{rapp_fct} in
the excitation region.

Finally, we can group the different terms of \eq{gammabeta} and \eq{C_S_ref}
in four sets:
\begin{eqnarray}
\label{S1}
S_1&=&\left|\deriv{ \xi_r}{r}\right|^2\,\approx\,\sigma^4\;
\left|\frac{\xi_r}{r}\right|^2\,,\\
\label{S2}
S_2&=& L^4 \left|\frac{\xi_h}{r}\right|^2\,
\approx\,\frac{\ell^4}{\sigma^4}\;
\left|\frac{\xi_r}{r}\right|^2\,,\\
\label{S3}
S_3&=&\left\{\,L^2\left|\deriv{ \xi_r}{r}\right|\left|\frac 
{\xi_h}{r}\right|,\,L^2\left|\deriv{ \xi_h}{r}\right|^2\, , L^2 \left| \deriv{\xi_h}{r}\right| \left|\frac{\xi_r}{r}\right| \right\} \nonumber \\
 &\approx&\,\ell^2\;\left|\frac{\xi_r}{r}\right|^2\,,\\
\label{S4}
S_4&=&\left\{\,\left|\frac{\xi_r}{r}\right|^2,\,\left|\frac{1}
{r}\deriv{|\xi_r|^2}{r}\right|,\,
\,L^2\left|\frac{\xi_r}{r}\right|\left|\frac{\xi_h}{r}\right|\,
\right\}\,.
\end{eqnarray}

The terms  in $S_4$ are always negligible compared to the others.
At fixed frequency (\,$\sigma$\,) we have thus:
\begin{alignat}{9}
\label{range1}
S_1 & \; \gg     \; & S_3 & \; \gg     \; & \; S_4 \; & \; \gg     \; & S_2 & \; \text{for } \ell \ll \sigma \\
\label{range2}
S_1 & \; \gg     \; & S_3 & \; \gg     \; & \; S_4 \; & \; \approx \; & S_2 & \; \text{for } \ell \approx \sigma \\
\label{range3}
S_1 & \; \approx \; & S_3 & \; \approx \; & \; S_2 \; & \; \gg     \; & S_4 & \; \text{for } \ell \approx \sigma^2 \\
\label{range4}
S_2 & \; \gg     \; & S_3 & \; \gg     \; & \; S_1 \; & \; \gg     \; & S_4 & \; \text{for } \ell \gg \sigma^2
\end{alignat}
In conclusion, the contribution of the horizontal displacement terms  ($S_2, S_3$)
begins to dominate the excitation for $\ell \gg \sigma^2$.


\subsection{Source terms as functions of $\ell$}

\emph{Reynolds stress contribution;}\\

We start by isolating non-radial effects in the range $\ell \in [0;500]$. 
Note that the limit $\ell=500$ is justified in Sect.~\ref{sep_ech} by the limit 
of validity of the present formalism. 
We investigate two cases, $\ell \ll \sigma^2$ and $\ell \approx \sigma^2$ respectively.
The condition for which  $\ell\simeq \sigma^2$
is satisfied for around the $f$ mode for  $\ell>50$ and 
in the gap between the $g_1$ and $f$ mode, for $\ell < 50$.

Using the set of inequalities \eq{range1} to \eq{range4}, for a typical
frequency of $3$ mHz (i.e. $\sigma \approx 30$), $R(r)$
(Eq.\,\ref{gammabeta}) becomes for \underline{high-$n$~modes} ($\ell \ll \sigma^2$):
\begin{eqnarray} 
\label{high_n} R (r) \approx \frac{16}{15} \left|
\deriv{\xi_r}{r} \right|^2 
\end{eqnarray} 
Hence, for high-$n$ acoustic
modes one can use \eq{high_n} instead of \eq{gammabeta}, and in terms
of the excitation source term, the formalism reduces to the radial case
for $\ell < 500$ and high-$n$ modes.

For \underline{low-$n$} modes ($\ell \approx \sigma^2$, i.e. for instance $\sigma \approx 10$) some additional dependency must be retained (see \eq{range3}). One gets
\begin{eqnarray}
\label{low_n}
R (r) &\approx& \ \frac{16}{15} \left| \deriv{\xi_r}{r} \right|^2 
+ \frac{16}{15} L^4 \left|\frac{\xi_h}{r}\right|^2  
- {2\over 5} L^2 \left(\deriv{\xi^*_r}{r} {\xi_h \over r} + c.c \right) \nonumber \\
&+& \frac{11}{15} L^2 \left| \zeta_r  \right|^2
\end{eqnarray}

The additional terms correspond to the non-diagonal contributions of the tensor $\vec \nabla : \vec \xi$ 
appearing in the Reynolds stress term $C_R^2$ because 
we are in the range $\ell \approx \sigma^2$ (see \eq{range3}).  
The radial and transverse components
of the divergence of the displacement nearly cancel so that
$\delta\rho/\rho$ takes its minimum values. 
This is due to the fact that they are nearly divergence free (minimum  
of $\delta \rho / \rho + \nabla \cdot \vec  \xi =0$), i.e.
\eqna{
\label{eqn:div1}
\vec \nabla_0 \cdot \,  \vec \xi = D_\ell \, Y_\ell^m \approx \deriv{\xi_r}{r} - L^2 \frac{\xi_h}{r} \approx 0
}
As the divergence of the mode corresponds to the diagonal part of the
tensor $\vec \nabla : \vec \xi$, one can then expect that the
excitation rate decreases (through the terms in $\textrm{d}\xi^*_r / \textrm{d}r  \times
\xi_h / r$ in \eq{low_n}).  However, such a decrease is compensated by
the non-radial component of the tensor ($\zeta_r ^2$ in \eq{low_n}). 
Thus, for low-$n$ $p$~modes there is a balance between the effect of
incompressibility that tends to diminish the efficiency of the
excitation and the non-diagonal components of the 
tensor $\vec \nabla : \vec \xi$ which tend to increase it. \\

\emph{Entropy contribution:}

Numerical investigations shows that the non-radial component of the entropy
source term does not significantly affect the excitation rates except
for $\ell > 1000$, which is out of the domain of validity of the
present formalism (see Sect.~\ref{sep_ech}).  The non-radial effects
appear through the mode compressibility, $L^2 \left| D_\ell
\right|^2$ (\eq{Bell}).  From \eq{range3} one can show that non-radial
contributions play a non-negligeable role for low-$n$ modes.  However,
such low-frequency modes are not sufficiently localized in the
superadiabatic zone, where  the entropy source term is maximum, to be
efficiently excited.

\section{Numerical estimations for a solar model}
\label{result}

\subsection{Computation of the theoretical excitation rates}
\label{calculation_method}

In the following, we compute the excitation rates of $p$~modes  for a solar model.
The rate ($P$) at which energy is injected into a mode per unit time is
calculated according to the set of 
Eqs.~(\ref{mean_square_amplitude_trapping}) - (\ref{eqn:C2S}). 
The calculation thus requires the knowledge of four different types of quantities:
\begin{itemize}
\item[1)] Quantities that are related to the
oscillation modes: the eigenfunctions ($\xi_r, \xi_h$) and associated
eigenfrequencies ($\omega_0$).
\item[2)] Quantities that are related to the spatial
and time-averaged properties of the medium: the density (\,$\rho_0$\,), 
the vertical velocity (\,$\tilde{w}$\,), the entropy (\,$\tilde s$\,), and  $\alpha_s=\partial P_1 / \partial \tilde{s}$.
\item[3)] Quantities that contain information about
spatial and temporal correlations of the convective fluctuations:
$E(k)$, $E_s(k)$, and $\chi_k(\omega)$.
\item[4)]  A quantity that takes anisotropy into account: $\Phi$ 
measures the anisotropy of the turbulence and 
is defined according to \cite{Gough77} (see also Paper~I for details) 
as:
\begin{equation}
\Phi=\frac{<u^2>}{<w^2>} \; ,
\end{equation}
where $u^2=w^2+u_h^2$ and $u_h$ is the horizontal velocity. \\
To be consistent with the current assumption of isotropic turbulence we 
assume $\Phi=3$.
\end{itemize}
Eigenfrequencies and eigenfunctions (in point 1) above) were computed
using the adiabatic pulsation code OSC \citep{boury75}. The solar
structure model used for these computations (quantities in point 2) was
obtained using the stellar evolution code CESAM \citep{Morel97} for the
interior, and a \cite{Kurucz93} model for the atmosphere.  The
interior-atmosphere match point was chosen at $\log\tau=0.1$ (above the
convective envelope).  The pulsation computations used the full
model (interior+ atmosphere).  In the interior model, we used the OPAL
opacities \citep{Opal96} extended to low temperatures with the
opacities of \cite{Alexander94}, the CEFF equation of state
\citep{CD1992}. Convection is included according  
to a B\"{o}hm-Vitense mixing-length formalism
\citep[see][for details]{Samadi06}, from which $\tilde{w}$ is
computed.  The $\Phi$ value is set to $2$ in the calculation.
This is not completely consistent  as we assume  isotropic 
turbulence (i.e. $\Phi=3$). This does not  however affect 
the conclusions of the present paper as
all results on nonradial excitation rates are normalized 
to the radial ones. Note also that the  equilibrium model does not 
include turbulent pressure.  These two limitations are of small importance
here as  our investigation 
in  this first work on nonradial  modes remains  essentially qualitative.

Finally, for the quantities in point 3, the total kinetic
energy contained in the turbulent kinetic spectrum ($E(k)$) is
 obtained following \cite{Samadi06}.

\begin{figure}[t]
\begin{center}
\includegraphics[height=6cm,width=9cm]{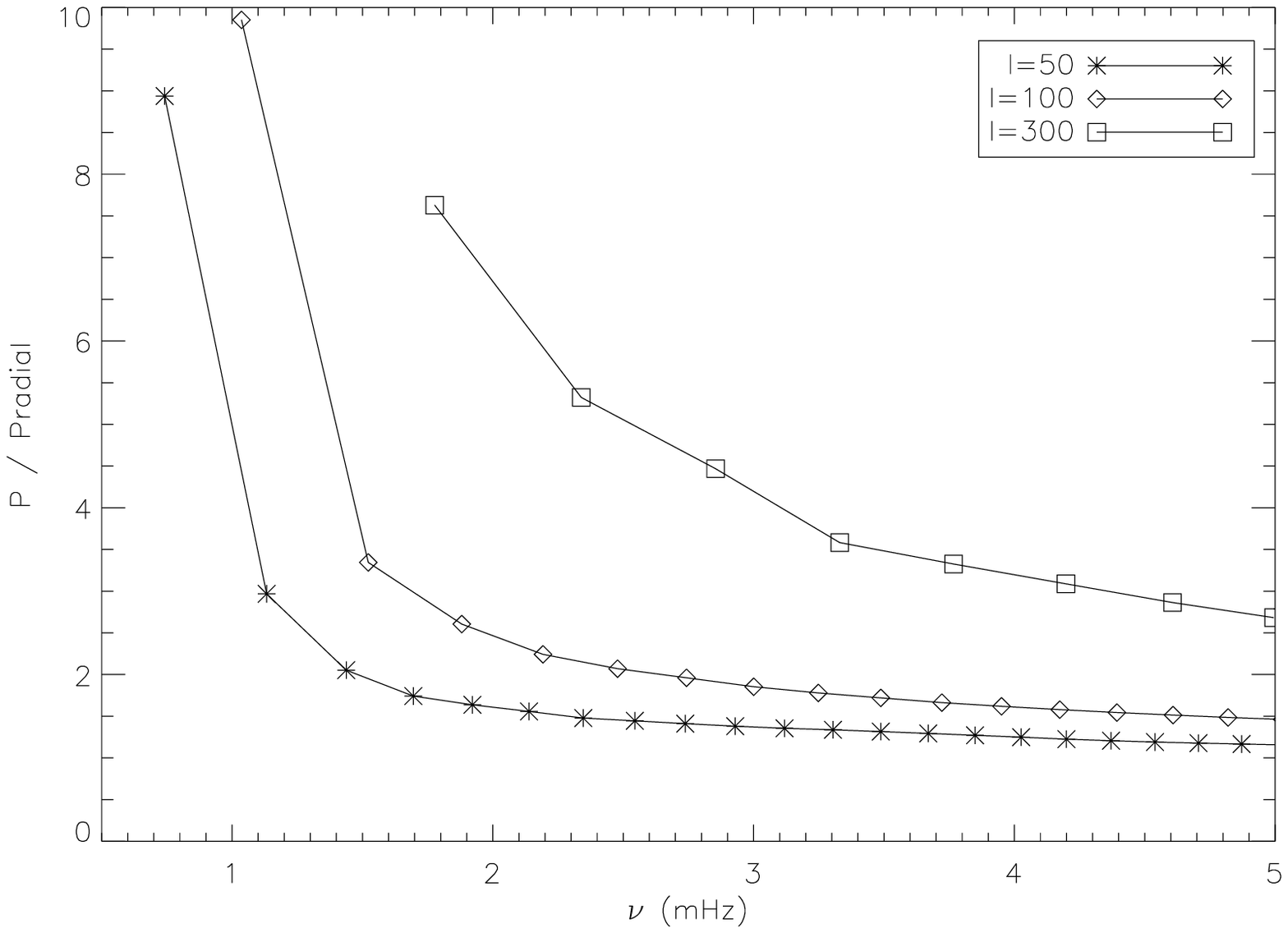}
\includegraphics[height=6cm,width=9cm]{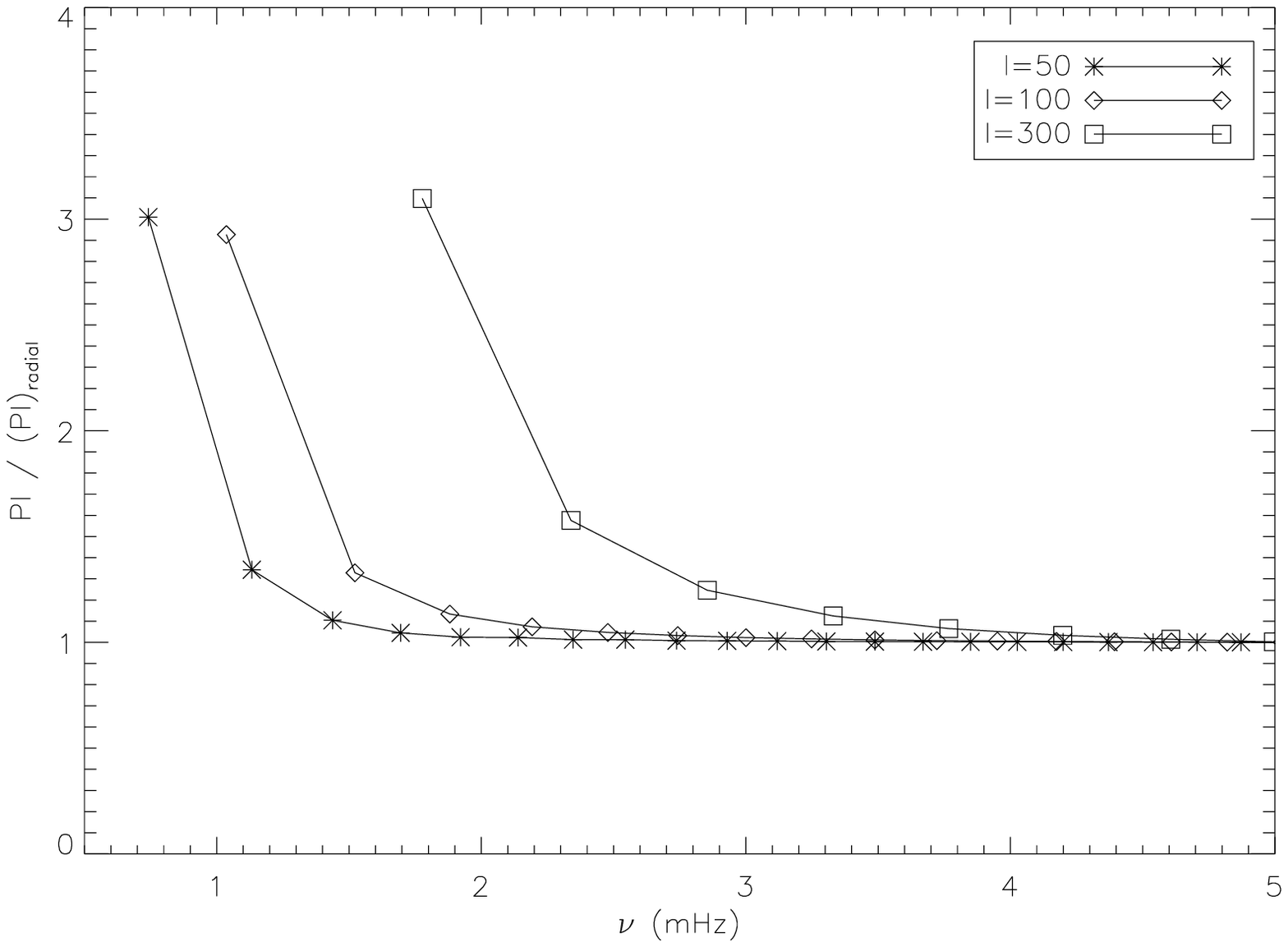}
\caption{{\bf top:} The rate (\,$P$\,) at which energy is supplied to each $\ell,n$ mode  for  $\ell=50, 100, 300$ is divided by
the excitation rate (\,$P_{radial}$\,)   obtained for  the $\ell=0, n$ mode. 
Computation of the theoretical excitation rates are performed as explained in 
Sect.~\ref{calculation_method}. {\bf bottom:}  Ratio $P~I/(P~I)_{radial}$ where $I$ is the
mode inertia. }
\label{p0}
\end{center}
\end{figure}

\subsection{ Excitation rates }
\label{Prates}

The rate (\,$P$\,) at which energy is supplied to the modes is plotted in Fig.~\ref{p0} normalized 
to the radial excitation rate (\,$P_{rad}$\,). 
It is seen that the higher the $\ell$, the more energy is supplied to the mode.  
This is explained by additional contributions (compared to the radial case)  due to 
mode inertia,  the spherical symmetry (departure from the plane-parallel assumption), 
and the contribution of horizontal excitation. 
Note that, as discussed in Sect.~\ref{radial} (see \eq{pp}), the departure from the plane-parallel 
approximation is negligible for $p$~modes. Then, 
to discuss the other two contributions, 
one can rewrite \eq{puissance_trapping} as 
\begin{equation}
\label{product}
P = \left( \frac{\left| \xi_r (R) \right|^2}{8 I} \right) \times \left( \frac{C_R^2 + C_S^2}{\left| \xi_r (R) \right|^2} \right)
\end{equation}
where $\left| \xi_r (R) \right|$ is taken at the photosphere. Note that
both terms of the product (\eq{product}) are independent of the
normalization of the eigenfunctions.  Thus, as shown by \eq{product},
the power supplied to the modes is composed of two contributions which
both depend on $\ell$:

The first is due to the mode inertia, which is defined as
\begin{equation} 
\label{inertia} I = \int_{0}^{M}
\textrm{d}m \, \left| \vec \xi \right|^2 = \int_{0}^{R} \,
\left( \left| \xi_r \right|^2 + L^2 \left| \xi_h \right|^2
\right) r^2 \, \rho_0 \textrm{d}r\;.  
\end{equation} 
High-$\ell$ modes
present a lower inertia despite the $L^2$ contribution in
\eq{inertia} because they are confined high in the atmosphere where the
density is lower than in deeper layers.

The second term of the product \eq{product} depends on the non-radial
effects through the excitation source terms (\eq{C_S_ref} and \eq{C2R_ref}). To investigate
this  quantity independent of the mode mass (defined as $I/\left| \xi_r (R) \right|^2 $), we plot the
ratio $PI / (PI)_{radial}$ in Fig.~\ref{p0}.
Excitation by horizontal turbulence plays a role at 
all values of $\ell$ and in the
same manner. One can then discuss two types of modes, namely
low-$n$ ($\leq 3$)and high-$n$ ($> 3$)  modes (see Fig.~\ref{p0}).
\begin{itemize}
\item For high-$n$ modes, non-radial effects play a minor role in the 
excitation source terms.
The dominant effect (see Fig.~\ref{p0}) is due to the mode inertia as
discussed above.
\item  For lower values of $n$,  there is a contribution to the
excitation rates due to the horizontal terms in \eq{C2R_ref}.
\end{itemize}
Thus, contrary to high-$n$ modes, the term $\deriv{\xi_r}{r}$ in $R(r)$
(\eq{gammabeta}) is no longer dominant in front of the terms involving 
$\xi_h$ for low-order modes.
Turbulence then supplies more energy to the  low-frequency
modes  due to horizontal contributions, which explains 
the higher excitation rates  for low-$n$ modes as seen  in Fig~\ref{p0}.
We stress that there is still turbulent energy that is supplied to the modes 
despite their nearly divergence-free nature. 
For such 
modes the non-diagonal part of the tensor $\vec \nabla : \vec \xi$, which is related 
to the shear of the mode, compensates 
and dominates the diagonal part, which is related to the mode compression.

\subsection{Surface velocities}
\label{surf}

Another quantity of interest is the theoretical surface velocity, 
which can be compared to observational data. 
We  compute the mean-squared surface velocity for each mode 
according to the relation \citep{Baudin05}:
\begin{equation}
\label{Pobs}
v_s^2 (\omega_0) = \frac{ P (\omega_0)}{2 \, \pi \,
\Gamma_\nu \, {\cal M} }
\end{equation}
where ${\cal M} \equiv I / \xi_{\rm r}^2(h) $ is the mode mass, $h$
the height above the photosphere where oscillations are measured,
$\Gamma_\nu = \eta/ \pi $ the mode linewidth at half maximum (in Hz),
and $v_s^2$ the mean square of the mode surface velocity.
Equation (\ref{Pobs}) involves the damping rates ($\Gamma_\nu$)  
inferred from observational data in the solar case for low-$\ell$ modes \citep[see][for details]{Baudin05}.  
We then assume  that the damping rates are roughly the same as for the 
$\ell=0$ modes. Such an assumption is supported for low-$\ell$ modes ($\ell \approx 50$) as shown 
by \cite{Barban04}. 

\begin{figure}[t!]
\begin{center}
\includegraphics[height=6cm,width=9cm]{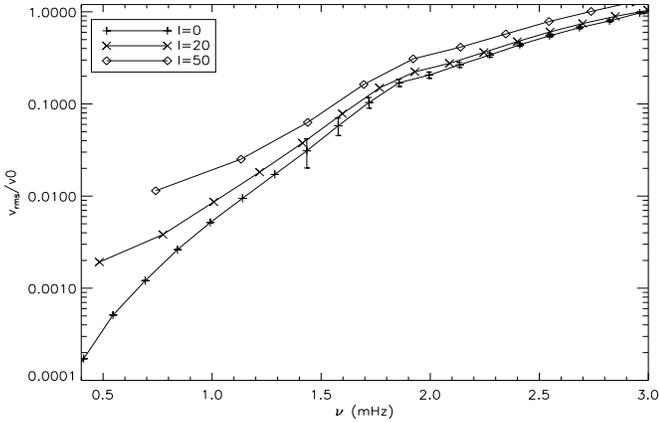}
\caption{Surface rms velocities of $\ell=0, 20, 50$ modes calculated using \eq{Pobs} and 
normalized to the maximum velocity of the radial modes (see text). Note that the 
damping rates are taken from GOLF \citep{Baudin05} 
and  are chosen to be the same for all angular degrees (\,$\ell$\,). 
Three $\sigma$ error bars derived from GOLF  
are plotted on the $\ell=0$ curve.}
\label{p3}
\end{center}
\end{figure}

Fig.~\ref{p3} displays the surface velocities for $\ell=0, 20$, and
$50$.  Note that the surface velocities are normalized to the
maximum velocity of the $\ell=0$ modes ($V_0 \approx$ 8.5~cm~s$^{-1}$
using MLT). This choice is motivated by the fact that the absolute
values of velocities depend on the convective model that is used and is
certainly imperfect.  However, its influence disappears when
considering differential effects.  As an indication, 3~$\sigma$ error
bars estimated from GOLF for the $\ell=0$ modes are plotted
\citep[see][for details]{Baudin05}.  
The differences between the radial
and non-radial computations are indeed larger than the $\ell=0$
uncertainties for $\ell > 20$.
For a more significant comparison, error bars for non-radial modes 
should be used but they are difficult to determine with confidence (work in progress).  
For $\ell$ larger than
$50$, we do not give surface velocities, as derived those here depend on the
assumption of approximately constant damping rate that is not confirmed for $\ell > 50$.

When available, observational data should allow us to
investigate the two regimes that have been emphasized in
Sect.~\ref{Prates}, namely the high- and low-$n$ modes.

\section{Discussion}
\label{discussion}

\subsection{The separation of scales}
\label{sep_ech}

\begin{figure}[t]
\begin{center}
\includegraphics[height=6cm,width=9cm]{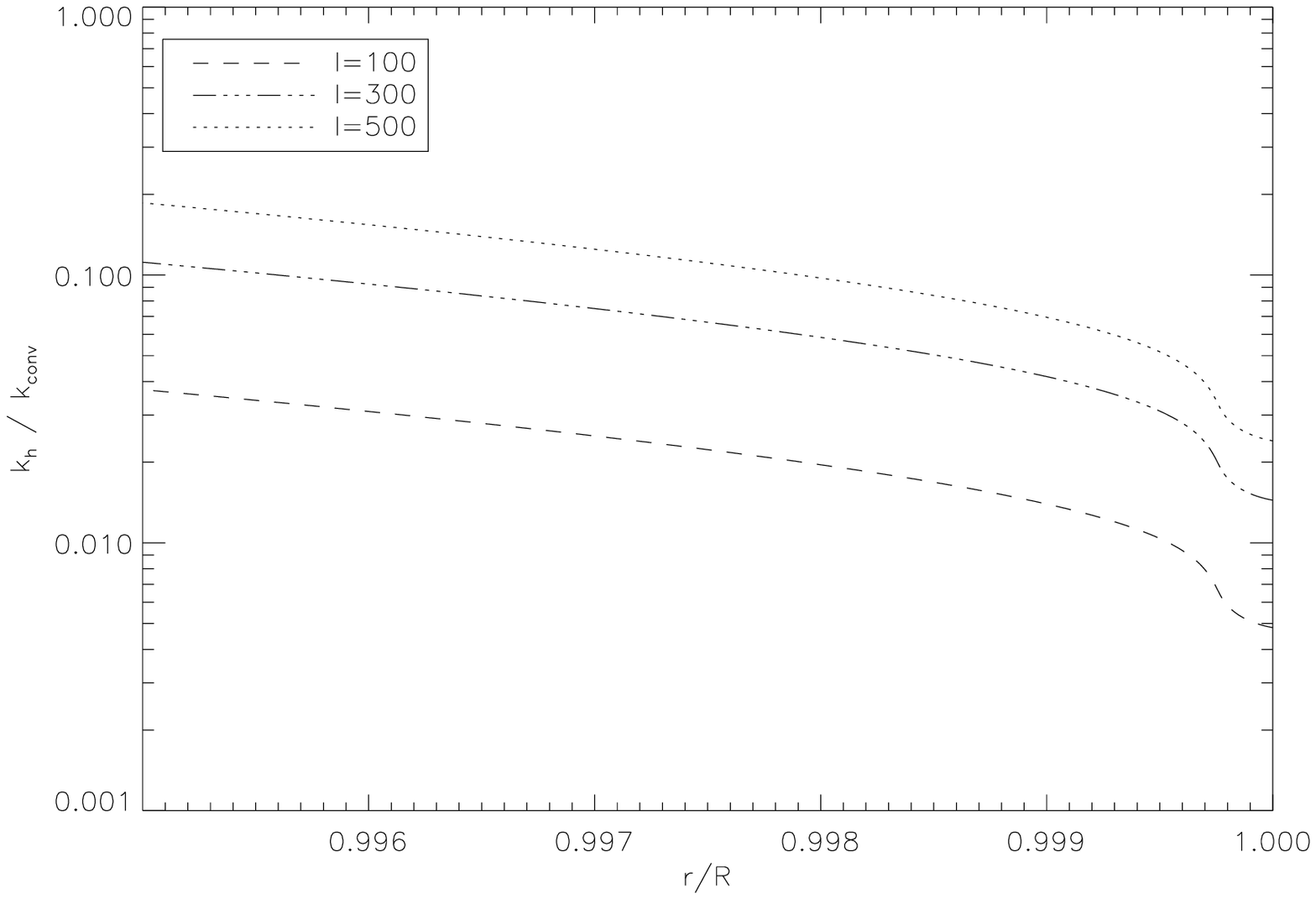}\\
\includegraphics[height=6cm,width=9cm]{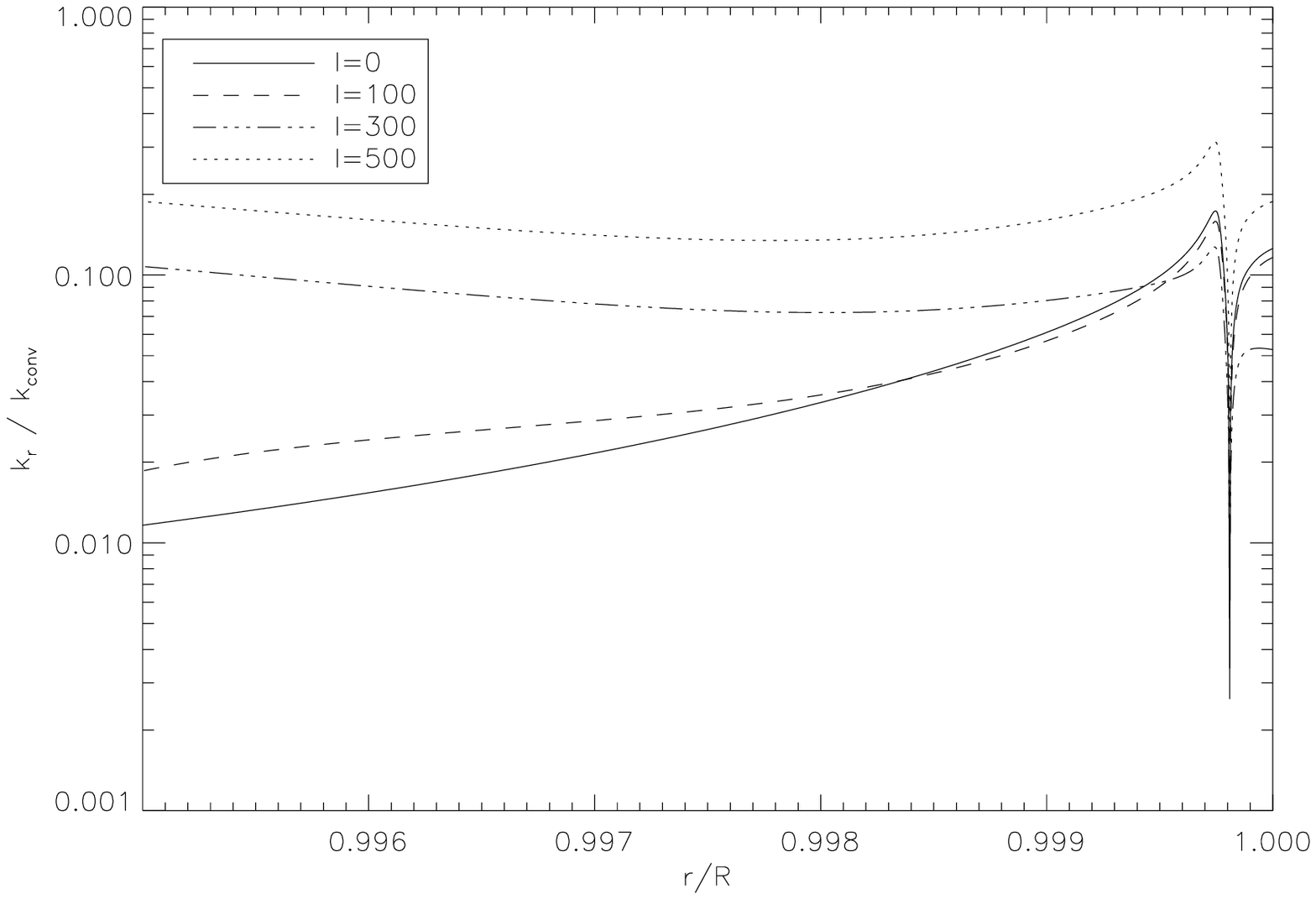}
\caption{{\bf Top:} Ratio of the horizontal oscillation wavenumber to the convective wavenumber 
($k_h / k_{conv}$), versus the normalized radius ($r/R$). $k_{conv}$ is computed using 
the mixing length theory such that $k_{conv} = 2 \pi / L_c$ ($L_c$ is 
the mixing length) and $k_{r}$ is computed using the dispersion relation \eq{dispersion}. 
Note that the ratio $k_h / k_{conv}$ is computed for a frequency of around $\nu= 3$ mHz, 
depending of the angular degree (\,$\ell$\,). 
{\bf Bottom:} the same as in the top but for the ratio $k_r / k_{conv}$.}
\label{sep_radial}
\end{center}
\end{figure}

The main assumption in this general formalism appears in \eq{mean_square_amplitude_trapping}, 
where it has been assumed that the spatial variation of the eigenfunctions is large compared 
to the typical length scale of turbulence, leading to what we call  \emph{the separation of scales}. 
In order to test this assumption, one must  compare the oscillation
wavelength to the turbulent one  or equivalently the wavenumbers. To this end, 
we use the dispersion relation \cite[see][]{Unno89}
\begin{equation}
\label{dispersion}
k_r^2 = \frac{\omega^2}{c_s^2}(1 - \frac{S_\ell^2}{\omega^2})(1 - \frac{N^2}{\omega^2})
\quad \mbox{ and} \quad k_h^2 = \frac{L^2}{r^2}
\end {equation}
where $N$ is the buoyancy frequency, $S_\ell$ the Lamb frequency and $k_r,k_h$ the radial and 
horizontal oscillation wavenumbers, respectively, and $L^2 = \ell(\ell+1)$.\\
For the turbulent wavenumber,  we choose to use, as a lower limit, the convective wavenumber 
$k_{conv} = 2 \pi / L_c$ where $L_c$ is the typical convective length scale. 
Thus, the assumption of separation of scales is fulfilled provided
\begin{equation}
k_{r,h} / k_{conv} \ll 1 
\end{equation}
In Figs.~\ref{sep_radial} are plotted the ratios $k_r/k_{conv}$ and
$k_h/k_{conv}$ respectively.  Those plots  focus on the uppermost part
of the solar convection zone where most of the excitation takes place.
The assumption of separation of scale  is valid for the horizontal
component of the oscillation since  one has $k_h / k_{conv} \ll 1$ (for
$\ell \leq 500$) in the region where excitation is dominant. However,
we must recall that our criterion  is based on  the mixing length to
compute $k_{conv}$.  As shown by \cite{Samadi02II} using 3D numerical
simulations, the convective length scale (computed using the CESAM
code, see Sect.~\ref{calculation_method}) must be multiplied by a
factor around five in order to reproduce the injection scale ($L_c$) in the
superadiabatic layers.  Hence, for a more conservative criterion, we
must then  multiply  the ratio $k_h / k_{conv}$ by a factor of five, this
leads to a ratio near unity for  $\ell \approx 500$ (see
Fig.~\ref{sep_radial}).  Thus, for higher values of the angular degree,
the separation-of-scale hypothesis becomes doubtful.

Concerning the radial component of the oscillation wavenumber, the
limiting value of $\ell$ seems to be the same (i.e. $\ell=500$).  
Thus, we conclude that for modes of angular degree lower than 500 one  
can use the separation of scales assumption. 
For $\ell >$~500
the characteristic length of the mode becomes smaller than the
characteristic length $L_c$ of the energy bearing eddies. Those
modes will then be excited by turbulent eddies with length-scale
smaller than $L_c$, i.e. lying in the turbulent cascade.  These
eddies inject less energy into the mode than the energy bearing eddies
do, since they have  less kinetic energy.  We can then expect that
-- at fixed frequency -- they received less energy from the turbulent eddies than the
low-degree  modes.  A theoretical development is currently underway
to properly treat the case of very high $\ell$ modes.

\subsection{The closure model}

A second approximation  in the present formalism is the use of a
closure model.  The uppermost part of the convection zone is a
turbulent convective system composed of two flows (upward and downward)
and the probability distribution function of the fluctuations of the
vertical velocity and temperature does not obey a Gaussian law
\citep{Lesieur97}.  Thus, the use of the quasi-normal approximation
(QNA, \citet{Million41}) which is exact for a normal distribution,
is no longer rigorously correct.  A more realistic  closure model has
been developed in \cite{Belkacem06a} and can be easily be adapted for
high-$\ell$ modes.  This alternative approach  takes into account  the
existence of two flows (the up- and downdrafts) within  the
convection zone.  However, the QNA is nevertheless often used for the sake
of simplicity as is the case here.  Note that, when using the
closure model with plumes, it is no longer  consistent to assume that
the third-order velocity moments strictly vanish, however as shown by
\cite{Belkacem06a,Belkacem06b}, their contribution is negligible in the
sense that their effect is smaller than the accuracy of the presently
available observational data.

\subsection{Mode inertia} 

We have shown that the excitation rates for
high-$\ell$ and $n$ modes are sensitive to the variation of the mode
inertia ($I$).  $I$ depends on the structure of the stellar model and
the properties of the eigenfunctions in  these external regions.
\cite{Samadi06} have shown that different local formulations of
convection can change the mode inertia by a small amount. This
sensitivity then affects the computed excitation rates ($P$).  However,
the changes induced in $P$  are found to be smaller than the accuracy to 
which the mode excitation rates  are derived from the current
observations \citep[see][]{Baudin05,Belkacem06b}.  Furthermore,
concerning the way the modes are obtained, we have computed
non-adiabatic eigenfunctions using the time-dependent formalism of
Gabriel for convection \citep[see][]{grigahcene05}.  The mode inertia
obtained with these non-adiabatic eigenfunctions exhibits a $\nu$
dependency 
different from those obtained using
adiabatic eigenfunctions (the  approximation adopted in the present
paper).  On the other hand, the mode inertia using non-adiabatic
eigenfunctions \citep[see][for details]{Houdek1999} obtained 
according to Gough's time-dependent formalism
of convection \citep{Gough77} shows smaller differences with the adiabatic
mode inertia.
Accordingly, the way the interaction of oscillation and time-dependent convection is
modelled affects the eigenfunctions differently.  As explained in
Sect.~\ref{surf}, the formalism developed in this paper can be an
efficient tool to derive constraints on the mode inertia to  
discriminate between the different treatments of convection.  Further
work is thus needed on that issue.

\section{Conclusions}

We have extended  the \cite{Samadi00I} formalism  in order 
to predict the amount of energy that is supplied to  non-radial modes.
In this paper, we focused on high-$\ell$ acoustic modes with a
particular emphasis on the solar case.  The validity of the
present formalism is limited to values of the angular degree lower than
$\ell=500$, due to the separation of scale assumption that is
discussed above, in Sect.~\ref{sep_ech}.  We have demonstrated that
non-radial effects are due to two contributions, namely the effect of
inertia that prevails for high-order modes ($n > 3$) and  non-radial
contributions in the Reynolds source term in $C_R^2$ (see \eq{C2R_ref}) 
that dominate over the radial one for low-order modes ($n < 3$).

 Contrary to \cite{Belkacem06b} who used 3D simulations to build an equilibrium 
model, we restrict ourselves to the use of a simple classical 1D MLT equilibrium model.
Indeed, we were interested in this work in deriving qualitative conclusions on nonradial 
contributions.
Forthcoming quantitative studies will have to use more realistic equilibrium models, 
particularly for the convection description, such as models including  turbulent pressure 
\citep[e.g.][]{B92} or 
patched models \citep[e.g.][]{Ros99}.  

From a theoretical point of view, several improvements and extensions
of the present formalism remain to be carried out.  For instance, one
must relax the assumption of the separation of scales if one wants to
use very high-$\ell$ modes. Such an investigation (which is currently
underway) should enable us  to  draw  conclusions about the
observational evidence that beyond some large value of $\ell$ the
energy supplied to the modes decreases with frequency
\citep[see][Fig.~2]{Woodard01}.  Another hypothesis is the isotropic 
turbulence that has been assumed in the present work as a first approximation. 
Such an assumption is to be given up to get a better description of nonradial 
excitation of modes by turbulent convection, which requires further 
theoretical developments (work in progress).

The present work  focuses on $p$~modes, but the formalism is valid for
both $p$ and $g$~modes.  We will address the analysis of gravity modes in a
forthcoming paper.

\newpage

\appendix
\section{Detailed expressions  for source terms}

The eigenfunctions (\,$\vec \xi$\,) are developed in spherical coordinates
$(\vec e_r, \vec e_\theta, \vec e_\phi)$
and expanded in spherical harmonics.
Hence the fluid displacement eigenfunction for 
a mode with given $n,\ell,m$ is written as
\eqn{
\label{decomp_harmonics}
\vec \xi (\vec r)= \left ( \xi_r \vec e_r + \, \xi_h \, \vec \nabla_H \right )
\, Y_{\ell,m} 
}
with
\eqn{
\vec \nabla_H =\left \{
\begin{array}{l}
\vec e_\theta \ds{ \derivp{}{\theta} } \vspace{0.3cm}  \\
\vec e_\phi \ds {\frac{1}{ \sin \theta} \derivp{}{\phi} }
\end{array}
\right .
}
where the spherical harmonics ($Y_{\ell,m} (\theta,\phi)$) are normalized according to 
\begin{equation}
\label{normalisation}
\int \frac{d \Omega}{ 4 \pi} \,  Y_{\ell,m} ~Y^*_{\ell,m}   = 1 
\end{equation}
with $\Omega$ being the solid angle ($\textrm{d}\Omega = sin \theta \textrm{d} \theta \textrm{d} \phi$).\\
The large-scale gradient $\vec \nabla_0$ appearing in \eq{eqn:C2S} and (\ref{eqn:C2Rb}) for instance 
is given, in the local spherical coordinates, by :
\eqn{
\vec \nabla_0 = \vec e_r  { \derivp{}{r} } + \frac{1}{r} \vec \nabla_H
\label{eqn:nalba0}
}

\subsection{Contribution of the turbulent Reynolds stress}
\label{C_R_detail}

The Reynolds stress contribution  can be written as (see Sect.~\ref{reynolds})
\begin{eqnarray}
C_R^2 & = &\pi^{2}   \int  {\textrm{d}^3 x_0}  \,
\left (\rho_0^2 \,  b^*_{ij} b_{lm} \right )
\int \textrm{d}^3k \, \int \textrm{d}\omega \,  \nonumber \\
&\times& \left (  T^{ijlm} + T^{ijml} \right )
\frac {E^2(k)} {k^4 }  \, \chi_k( \omega_0+\omega) \, \chi_k(\omega)
\label{eqn:C2R_2}
\end{eqnarray}
where
\eqn{
\label{Tijlm}
T^{ijlm} = \left( \delta^{il}- \frac {k^i k^l} {k^2}  \right)   \left( \delta_{jm}-
\frac {k^j k^m} {k^2}  \right) \; .
}
and
\eqn{
\label{eqn:bij}
b_{ij} \equiv \vec {e}_i \, \cdot \, \left ( \vec \nabla_0 \, : \, \vec \xi \right ) \, \cdot
\,  \vec {e}_j
}
where the double dot denotes the tensor product.

We now consider the covariant $(\vec a^r , \vec a^\theta,\vec a^\phi)$ and the
contravariant $(\vec a_r , \vec a_\theta,\vec a_\phi)$ natural base coordinates 
where the eigenfunction can be expanded 
\eqn{
\label{composantes}
\vec \xi 
=  \hat \xi^k \vec e_k  = q_k \vec a^k  ~~~~~~~k=\{ r,\theta,\phi \}
}
The natural and physical coordinates are related to each other by 
\eqn{
\vec e_i = \frac{1}{\sqrt{ | g_{ii} | }} \vec a_i
\label{eqn:ui_ei}
}
where $g_{ij}$ is the metric tensor  in spherical coordinates \citep[see Table~6.5-1 in ][]{Korn68}, i.e.
\begin{eqnarray}
g_{rr} &=& 1 \, , g_{\theta\theta} = r^2 \, ,  g_{\phi\phi} = r^2 \sin^2 \theta \, , \nonumber \\ 
g_{ij} &=& 0  \mbox{    for }  i \neq j \; .
\end{eqnarray}

Equation~(\ref{eqn:bij}), with the help of \eq{composantes}, 
can then be developed in  covariant coordinates
\eqna{
\vec \nabla_0 \,:\, \vec \xi = \vec a^i \, \derivp{\vec \xi}{x^i} \,  = \vec a^i \,
\vec a^j \, \left ( \derivp{q_j}{x^i } \right ) + \vec a^i \, q_j \,
\left ( \derivp{\vec a^j}{x ^i}  \right )
\label{eqn:nabal0_xi} \nonumber \\
= \vec a^i\,  \vec a^j\,  \left ( \derivp{q_j}{x^i} \right ) - q_j \,
\Gamma_{pi}^j \,  \vec a^i \,  \vec a^p
}
where  $ \Gamma_{pi}^j$ is the Christoffel three-index symbol of the second kind \citep{Korn68}.
According to \eq{eqn:nabal0_xi} and \eq{eqn:ui_ei}, 
$b_{ij}$ (\eq{eqn:bij}) can be written as:
\eqn{
b_{ij} = \frac{1}{\sqrt{| g_{ii} \, g_{jj} | } } \left ( \derivp{q_j}{x^i } -
q_p \,   \Gamma_{ji}^p \right )
\label{eqn:bij_2}
}
To proceed, one has to express \eq{eqn:bij_2} in terms of the physical coordinates ($\hat \xi^k$). 
With the help of \eq{composantes} and \eq{eqn:ui_ei}, we relate  the covariant coordinates $q_i$
to the physical (contravariant) components $\hat \xi^j$ 
\eqn{
q_j  = \frac{g_{ij}} { \sqrt{ | g_{jj}| }} \,  \hat \xi^j  
\label{eqn:xi_hatxi}
}
where the component $\hat \xi^k$ are derived from \eq{decomp_harmonics}
\eqn{
\hat \xi^r = \xi_r Y_{\ell,m} ;\\
\hat \xi^\theta = \ds{ \xi_h\derivp{Y_{\ell,m} }{\theta} } ; \\
\hat \xi^\phi =  \xi_h \ds{  \frac{1}{\sin \theta} \derivp{Y_{\ell,m} }{\phi} }\\
}
Hence \eq{eqn:bij_2} becomes:
\eqn{
\begin{array}{lll}
\vspace{0.2cm} b_{rr} & = &  \ds \left ( \deriv{\xi_r }{r} \right) \, Y_{\ell,m}   \nonumber \\
\vspace{0.2cm} b_{r \theta}  & =  & \ds \left(  \deriv{\xi_h}{r}  \right)  \,
\derivp{ Y_{\ell,m}}{\theta}      \nonumber  \\
\vspace{0.2cm} b_{r \phi}  & =  & \ds    \left(   \deriv{\xi_h}{r}
  \right) \frac{1}{\sin \theta }
\,  \derivp{Y_{\ell,m} }{\phi}        \nonumber \\
\vspace{0.2cm} b_{\theta r} & = & \ds  \frac{1}{r} (\xi_r -  \xi_h)
 \derivp{Y_{\ell,m}}{\theta}   \nonumber \\
\vspace{0.2cm}  b_{\theta \theta} & = & \ds  \frac{\xi_h}{r}
\left ( \derivp{^2  Y_{\ell,m}}{\theta^2} \right ) +  \frac{\xi_r}{r} Y_{\ell,m}     \nonumber \\
\vspace{0.2cm}  b_{\theta \phi} & = &  b_{\phi \theta }= \ds\frac{ \xi_h}{r} 
\derivp{}{\theta}\left [\frac{1}{\sin \theta }
\derivp{Y_{\ell,m}}{\phi} \right ]
  \nonumber \\
\vspace{0.2cm}  b_{\phi r} & = & \ds  \frac{1}{r} (\xi_r - \xi_h) \frac{1}{ \sin \theta }
\derivp{Y_{\ell,m}}{\phi}    \nonumber \\
\vspace{0.2cm}  b_{\phi \phi} & = & \ds \frac{\xi_r}{r} Y_{\ell,m}  + \frac{\xi_h}{r} 
  \ds \left [ \frac{1}{\sin^2 \theta }\left ( \derivp{^2 Y_{\ell,m}}{\phi^2}  \right )
+   \frac{\cos \theta}{\sin \theta} ~
\left ( \derivp{Y_{\ell,m}}{\theta} \right ) \right ]
\label{eqn:bij_4}
\end{array}
}
The contribution of the Reynolds stress can thus be written as:
\begin{eqnarray}
C_R^2 & =  &  4\pi^{3}   \int  \textrm{d}m  \,\int \textrm{d}k ~\int \textrm{d}\omega \,  \; R(r, k)
 \nonumber \\  &\times& \,
\frac {E^2(k)} {k^2 }  \chi_k( \omega + \omega_0) \chi_k( \omega )  \; ,
\label{eqn:C2R_3}
 \end{eqnarray}
where we have defined $\textrm{d}m= 4\pi r^2 \rho_0 \textrm{d}r$ and 
\begin{eqnarray}
R (r, k) &\equiv& \int {\textrm{d}\Omega\over 4 \pi} ~ \int  {\textrm{d}\Omega_k \over 4 \pi}\, b^*_{ij}
\, b_{lm} \, \left ( T^{ijlm} + T^{ijml} \right ) 
\end{eqnarray}
Because $T^{ijlm}=T^{jiml}$, it is easy to show that 
\begin{eqnarray}
R (r, k) &\equiv& \int {\textrm{d}\Omega\over 4 \pi} ~ \int  {\textrm{d}\Omega_k \over 4 \pi} \, B^*_{ij}\, B_{lm} \, 
\left ( T^{ijlm} + T^{ijml} \right ) \nonumber 
\end{eqnarray}
where $B_{ij}\equiv  (1/2)(b_{ij}+b_{ji})$. 

Using the expression \eq{Tijlm} for $T^{ijlm}$, we  write 
\begin{eqnarray}
\label{R}
R (r, k) &\equiv & R_1- R_2+R_3 
\end{eqnarray}
where
\begin{eqnarray}
& &R_1  =   2   \int {\textrm{d}\Omega\over 4 \pi} ~ \int  {\textrm{d}\Omega_k \over 4 \pi} \,   \left( \sum_{i,j}~ B^*_{ij} B_{ij} 
\right)  \nonumber\\
& & R_2 =  4  \int {\textrm{d}\Omega\over 4 \pi} ~ \int  {\textrm{d}\Omega_k \over 4 \pi} \, 
\left(\sum_{i,j} ~ B^*_{ij} B_{il} \frac{k_j k_l}{k^2} \right) \nonumber \\
& & R_3 =   2 ~\int {\textrm{d}\Omega\over 4 \pi} ~ \int  {\textrm{d}\Omega_k \over 4 \pi} \,  
\left(\sum_{i,j} B^*_{ij} B_{lm} \frac{k_i k_j k_l k_m}{k^4}  \right) \, .
\label{R12}
\end{eqnarray}

We assume isotropic turbulence, hence the $\vec k$  components satisfy
$$ \int \textrm{d}\Omega_k ~\frac{k_i k_j}{k^2}  = \delta_{ij} \int \textrm{d}\Omega_k ~  \frac{k^2_r}{k^2}$$
with $\delta_{ij}$ is the Kronecker symbol for $i,j=r,\theta,\phi$. Then we obtain
\begin{eqnarray}
R_1  &=&   2  \int {\textrm{d}\Omega\over 4 \pi} ~    \left( \sum_{i,j}~ \vert B_{ij} \vert^2  \right)
\nonumber
\\
R_2  &=& 4  \int {\textrm{d}\Omega\over 4 \pi} ~ \int {\textrm{d}\Omega_k \over 4 \pi} ~\frac{k_r^2}{k^2}  ~ (\sum_{i,j} \vert B_{ij}
\vert^2)  =  \ 2  \alpha ~R_1 \label{R2} \nonumber \\
R_3  &=&  2 \beta ~\int {\textrm{d}\Omega\over 4\pi} ~ \left( \sum_{i,j} \vert B_{ij} \vert^2 
+  \sum_{i \not=j}  (B^*_{ii} B_{jj}+cc)  \right) \nonumber \\
  &=&  \beta  ~R_1  + 2 ~\beta  ~  \left( \int {\textrm{d}\Omega\over 4\pi} 
  ~\sum_{i \not=j} \left(B^*_{ii} B_{jj} + c.c \right)  \right) 
\label{R3}
\end{eqnarray}
where we have set 
\begin{equation}
 \alpha \equiv  \int {\textrm{d}\Omega_k \over 4 \pi} ~  \frac{k_r^2}{k^2}  ~; ~
  \beta \equiv \int {\textrm{d}\Omega_k \over 4 \pi} ~    \frac{k_r^4}{k^4} \end{equation} 
To compute $R_1,R_2$ and $R_3$, we write
\begin{eqnarray}
\label{eqB}
B_{rr}&=& b_{rr}; B_{\theta\theta}= b_{\theta\theta} ; B_{\phi\phi}=b_{\phi\phi}  \nonumber\\
B_{r\theta} &=&  \frac{1}{2} ~\zeta_r
\derivp{Y_{\ell,m}}{\theta}\nonumber \\
B_{r\phi} &=& \frac{1}{2} ~\zeta_r  \frac{1}{\sin\theta}  
\derivp{Y_{\ell,m}}{\phi} \\
B_{\theta \phi} &=& b_{\theta\phi}=b_{\phi\theta} \nonumber 
\end{eqnarray}
with 
\begin{equation}
\zeta_r =  \deriv{\xi_h}{r} +\frac{1}{r}(\xi_r-\xi_h)  
\end{equation}

Using the expression \eq{eqB} for the quantities $B_{ij}$,  
we obtain  after some manipulation:
\begin{eqnarray}
\label{R1}
R_{1}  &=& 2 \left| \deriv{\xi_r}{r} \right|^2 +  4 \left| \frac{\xi_r}{r}  \right|^2  + 2 L^2 (L^2 - 1) \left| \frac{\xi_h}{r}\right|^2 \nonumber \\
&+&  L^2 \left( \left| \zeta_r  \right|^2 - 2 (\frac{\xi_r^* \xi_h}{r^2} + c.c) \right)
\end{eqnarray}
For $R_3$, some lengthy manipulation leads to:
\begin{eqnarray}
\label{R3}
R_{3} / \beta &=& 2 \left| \deriv{\xi_r}{r} \right|^2 +  8 \left| \frac{\xi_r}{r}  \right|^2  +  2 \left(L^4 + 4 {\cal F}_{\ell,|m|}\right) \left| \frac{\xi_h}{r}\right|^2 \nonumber \\
&+&  L^2 \left( 2 \left| \zeta_r  \right|^2 - 2 (\deriv{\xi^*_r}{r} \frac{\xi_h}{r} + 2 \frac{\xi_r^* \xi_h}{r^2} + c.c) \right) \nonumber \\
&+& 4 (\frac{\xi^*_r}{r} \deriv{\xi_r}{r} + c.c )
\end{eqnarray}
where we have defined 
\begin{eqnarray}
{\cal F}_{\ell,|m|} &=&  \frac{(2 \ell + 1)}{2} \left( (|m|+1) A_{\ell,|m|}^2 + (|m|-1) B_{\ell,|m|}^2 \right) \\
A_{\ell,m}^2 &=& \frac{1}{4} \left( \ell (\ell + 1) - m (m+1) \right) \\
B_{\ell,m}^2 &=& \frac{1}{4} \left( \ell (\ell + 1) - m (m-1)  \right)
\end{eqnarray}
To derive $R_1, R_2$, and $R_3$ the following 
relations have been used 
\begin{equation}
\derivp{^2 Y_{\ell,m}}{\theta^2} + \frac{\cos \theta}{\sin \theta} \; \derivp{Y_{\ell,m}}{\theta} + \frac{1}{\sin^2 \theta} \derivp{^2 Y_{\ell,m}}{\phi^2}  = - L^2 Y_{\ell,m}
\end{equation}
\begin{equation}
-m \frac{\cos \theta}{sin \theta} Y_{\ell,m} = A_{\ell,m} Y_{\ell,m+1} e^{-i \phi} + B_{\ell,m} Y_{\ell,m-1} e^{i \phi}
\end{equation}
\begin{equation}
\derivp{Y_{\ell,m}}{\theta} = A_{\ell,m} Y_{\ell,m+1} e^{-i \phi} - B_{\ell,m} Y_{\ell,m-1} e^{i \phi}
\end{equation}
\begin{eqnarray}
& &\int  {\textrm{d}\Omega\over 4 \pi} ~ \left(\nabla_H Y^*_{\ell,m} \cdot  \nabla_H Y_{\ell,m}\right) =  L^2 \\
- & &\int  {\textrm{d}\Omega\over 4 \pi} ~ \left(\nabla^2_H Y^*_{\ell,m}\right)  Y_{\ell,m} = L^2 \\
& & \int  {\textrm{d}\Omega\over 4 \pi} ~  \left| \partial_\theta \left(\frac{1}{sin \theta} \derivp{Y_{\ell,m}}{\phi} \right)  \right|^2 = {\cal F}_{\ell,|m|}
\end{eqnarray}

Combining \eq{R1}, \eq{R2}, \eq{R3}, and \eq{R}, with  $\alpha  = 1/3,  \beta  =  1/5$ (isotropic turbulence, see Paper~I for details), yields
\begin{eqnarray}
\label{Rfinal}
R (r) &=&   {16\over 15} ~  \left| \deriv{\xi_r}{r}  \right|^2  +  {44\over 15} ~    \left| \frac{\xi_r}{r}  \right|^2
+   \frac{4}{5} \frac{1}{r} \deriv{\vert \xi_r\vert^2}{r} \nonumber \\ 
&+&  ~ L^2 \left( {11\over 15} ~  \left| \zeta_r  \right|^2 - {22 \over 15} (\frac{\xi_r^* \xi_h}{r^2} +c.c) \right) \nonumber \\
 &-&{2\over 5} L^2 \left(\deriv{\xi^*_r}{r} {\xi_h \over r} 
+ c.c \right)\nonumber \\
&+&  \left| \frac{\xi_h}{r}  \right|^2 
\left( \frac{16}{15} L^4+ \frac{8}{5} {\cal F}_{\ell,m} - \frac{2}{3} L^2 \right) 
\end{eqnarray}

For radial modes, this reduces to :
\begin{eqnarray}
R (r) &=&   {16\over 15} ~ \left| \deriv{\xi_r}{r} \right|^2  +   {44\over 15} ~  \left| \frac{\xi_r}{r} \right|^2
 +\ {4\over 5}~ \deriv{\left| \xi_r \right|^2}{r}   
\end{eqnarray}

The final expression for the Reynolds source term is then given by:
\begin{eqnarray}
C_R^2 & =  &  4\pi^{3}   \int  \textrm{d}m   \,  \; R(r)~ S_R(\omega_0)\; ,
\label{eqn:C2R_3}
 \end{eqnarray}
with
\begin{eqnarray}
S_R(\omega_0) = \,\int  \frac {\textrm{d}k} {k^2 }~E^2(k) ~\int \textrm{d}\omega ~\chi_k( \omega + \omega_0) \chi_k( \omega )
 \end{eqnarray}
and $R(r)$ by \eq{Rfinal}.

\subsection{Contribution of entropy fluctuations}
\label{C_S_detail}
We start from \eq{C2S_62}, and  to proceed further in the derivation of the entropy
fluctuation source term one has to compute
\begin{eqnarray}
& &\int \textrm{d} \Omega_k \,  h^{ij}  \,    T_{ij} 
\label{eqn:int_hT_over_omegak}
\end{eqnarray}
Then, $\vec \xi$ and $\vec k$ are expanded in
spherical coorindates $(\vec a_r, \vec a_\theta, \vec a_\phi)$.  
We assume an isotropic  turbulence, as a consequence the quantities $k_r \,
k_\theta$, $k_r k_\phi$, $k_\theta k_\phi$ vanish when integrated
over  $\Omega_k$. One next obtains
\begin{eqnarray}
\int \textrm{d} \Omega_k \,   h^{ij} \,  T_{ij} 
 = 2 \pi \, \mathcal{H} \, \left ( h_{rr}+ h_{\theta \theta} + h_{\phi \phi} \right )
\label{eqn:int_hT_over_omegak2}
\end{eqnarray}
where $\mathcal{H}$ is the anisotropy factor introduced in Paper~I which in the
current assumption (isotropic turbulence)  is equal to  $ 4/3 $.
Assuming that $\alpha_s=\alpha_s(r)$ we have according to  Eq.\,(\ref{eqn:nalba0})
\eqn{
\left \{
\begin{array}{lcl}
h_{rr} & = & \ds{ \left | ~ \mathcal{C} \,   \deriv{ \ln \mid \alpha_s \mid } {r}
-   \derivp{ \mathcal{C} } {r} \right | ^2 } \vspace{0.3cm}\\
h_{\theta \theta} & =  &\ds{  \frac{1}{r^2} \, \left |
\derivp{\mathcal{C}}{\theta}\right | ^2 } \vspace{0.3cm} \\
h_{\phi \phi} & = &\ds{  \frac{1}{r^2\, \sin^2 \theta } \,
\left | \derivp{\mathcal{C}}{\phi}\right | ^2 }
\end{array}
\label{eqn:h_components}
\right .
}

To proceed, it is necessary to express the divergence of the
eigenfunction
\eqna{ \mathcal{C} \equiv \vec \nabla_0
\cdot \,  \vec \xi = D_\ell \, Y_\ell^m \label{eqn:div_xi} 
} 

with 
\eqna{
 D_\ell(r,\ell) \equiv  D_r - \frac{L^2 } {r} \, \xi_h ~~;~~
D_r \equiv \frac{1}{r^2} \, \derivp{}{r} \left (  r^2  \xi_r \right ) \; ,
}
where  again  $L^2= \ell(\ell+1)$.\\
We next integrate Eq.\,(\ref{C2S_62})  over $\textrm{d} \Omega/ 4\pi$, the solid
angle associated with the eigenfunctions $ \vec \xi$.
One obtains with the help of Eq.\,(\ref{eqn:div_xi}) and according to Eq.\,(\ref{eqn:h_components})
\begin{eqnarray}
& & \int \frac{ \textrm{d}  \Omega}{4 \pi} \,  \int \textrm{d}  \Omega_k \, h^{ij} \, T_{ij} = \nonumber \\
& &\frac{ 2 \pi \mathcal{H}}{r^2} \,  \left (L^2 \, \left | D_\ell \right | ^2
+  \left |  D_\ell \, \deriv{\ln \mid \alpha_s \mid}{\ln r}  - \deriv{D_\ell}{\ln r}
 \right |  ^2   \right )
\label{eqn:hij_Tij}
\end{eqnarray}

The final expression for the contribution of entropy fluctuations reduces to:
\eqn{
\label{C_S_ref}
C_S^2  = \frac{4 \pi^3 \, \mathcal{H}}{\omega_0^2}  \, 
\int \textrm{d}^3 x_0 \, \alpha_s^2 \, \left ( A_\ell + B_\ell \right ) \,
\mathcal{S}_S(\omega_0)
}
where $\mathcal{H}$ is the anisotropy factor introduced in Paper~I that
 in the current
assumption (isotropic turbulence)  is equal to  $ 4/3 $.  In addition,
\eqna{
\label{C_S_ref2}
A_\ell & \equiv & \frac{1 }{r^2}  \,  \left|  D_\ell \, 
\deriv{\left( \ln \mid \alpha_s \mid \right)}{\ln r}
 - \deriv{D_\ell }{\ln r}  \right|  ^2    \label{eqn:Al}
\\
\label{Bell}
B_\ell & \equiv &  \frac{1 }{r^2} \, L^2 \, \left| D_\ell \right| ^2
\label{eqn:Bl}
\\
\mathcal{S}_S(\omega_0) & \equiv & \int \frac{\textrm{d}k}{k^4}\,E(k)
\, E_s(k) \nonumber \\ 
& &\hspace{+1cm} \times \int \textrm{d}\omega \,
\chi_k(\omega_0+\omega)\,  \chi_k(\omega)
\label{eqn:FS} 
} 

\end{document}